\documentstyle[preprint,tighten,pra,aps,latexsym,amssymb,epsfig]{revtex} 

\newcommand{\bdi}{\begin{displaymath}}
\newcommand{\edi}{\end{displaymath}}
\newcommand{\bfi}{\begin{figure}}
\newcommand{\efi}{\end{figure}}

\newcommand{\beq}{\begin{equation}}
\newcommand{\eeq}{\end{equation}}
\newcommand{\beqa}{\begin{eqnarray}}
\newcommand{\eeqa}{\end{eqnarray}}
\newcommand{\no}{\nonumber}

\newcommand{\rmd}{{\mathrm d}}
\newcommand{\msmall}{m}                               
\newcommand{\cmb}{cos\-mic mi\-cro\-wave back\-ground}
\newcommand{\CS} {Chern--Si\-mons}
\newcommand{\MCS}{Max\-well--Chern--Si\-mons}
\newcommand {\bcs}    {boundary conditions}

\newcommand {\qfth}   {quantum field theory}

\newcommand {\YM}     {Yang--Mills}

\newcommand {\rhs}    {right-hand side}
\newcommand {\SM}     {Standard Model}
\begin{document} \draft


\title{\LARGE \bf CPT violation: mechanism and phenomenology\thanks{to
       $\!$appear $\!$in:  $\!$\emph{Proceedings $\!$7-th $\!$International
       $\!$Wigner $\!$Symposium}, $\!$edited $\!$by $\!$Y.S. Kim $\!$and $\!$M.E. Noz.}
       }

\author{Frans R. Klinkhamer\thanks{Email address: frans.klinkhamer@physik.uni-karlsruhe.de
        \hspace*{\fill} KA--TP--28--2001, hep-th/0110135
        }}
\address{Institut f\"ur Theoretische Physik, Universit\"at Karlsruhe,
         D--76128 Karlsruhe, Germany}

\maketitle

\begin{abstract}
A new mechanism for T and CPT violation is reviewed, which relies on
chiral fermions, gauge interactions and nontrivial spacetime topology.
Also  discussed are the possible effects on the propagation of
electromagnetic waves \emph{in vacuo},
in particular for the cosmic microwave background radiation.
\end{abstract}


\section{Introduction}

The CPT theorem \cite{L54,P55,B55,L57,J57}
states that any local relativistic \qfth~is invariant under the combined
operation of charge conjugation (C), parity reflection (P) and
time reversal (T). Very briefly, the main inputs are (cf. Ref. \cite{L57}):
\begin{itemize}
\item the Minkowski spacetime;
\item the invariance under transformations of the proper
      or\-tho\-chro\-nous Lorentz group ${\mathcal L}_+^\uparrow$ and
      spacetime translations;
\item the normal spin-statistics connection;
\item the locality and hermiticity of the Hamiltonian.
\end{itemize}
An extensive discussion of this theorem and its consequences
can be found in Refs. \cite{BLOT90,S64}.

Here, we go further and ask the following question:
\emph{can} CPT invariance be violated at all
in a physical theory and, if so, \emph{is} it in the real world?
It is clear that something  ``unusual''  is required for this to be the case.
Two possibilities, in particular, have been discussed in the literature.

First, there is \emph{quantum gravity}, which may or may not lead to
CPT violation; cf. Ref. \cite{W80}.
The point is, of course, that Poincar\'{e} invariance does not hold in
general. Still, a CPT theorem can be ``proven''
in the Euclidean  formulation for asymptotically-flat spacetimes \cite{AG83}.
In the canonical formulation, on the other hand, there is an indication that certain
semiclassical (weave) states could affect the Lorentz invariance of
Maxwell theo\-ry at the Planck scale and break  CPT invariance \cite{GP99}.

Second, there is \emph{superstring theory}, which may or may not lead to CPT violation;
cf. Ref. \cite{S89}. The point, now, is the (mild) nonlocality of the theory.
There exists, however, no convincing calculation showing the necessary violation of CPT.

In this talk, we discuss a third possibility:
\emph{for certain spacetime topologies and classes of
chiral gauge theories, Lorentz and CPT invariance are broken by quantum effects.}
The main paper on this ``CPT anomaly'' is Ref. \cite{K00}.
(The relation with earlier work on sphalerons, spectral flow, and anomalies,
is explained in Ref. \cite{K98}.)
Further aspects of the CPT anomaly have been considered in
Refs. \cite{KN01,KM01,AK01npb,AK01plb} and a brief review has already appeared in Ref.
\cite{K01hep2001}.

The outline of the present write-up is as follows.
In Sec. II, a realistic (?) example of the CPT anomaly is given and some
general properties are emphasized.
In Sec. III,  the CPT anomaly is established for a class of exactly
solvable two-dimensional models (specifics are relegated to Appendix A).
In Sec. IV, the CPT anomaly is obtained perturbatively for a class of
four-dimensional chiral gauge theories, which includes the example of Sec. II.
The resulting modification of Maxwell theory is briefly discussed (the important
issue of microcausality is dealt with in Appendix B).
In Sec. V, certain phenomenological aspects of  the modified Maxwell theory
are mentioned, in particular as regards the propagation of light over large
distances (e.g., the \cmb~radiation).
In Sec. VI, finally, some conclusions are drawn.

\section{EXAMPLE AND GENERAL REMARKS}

The CPT anomaly is best illustrated by  a concrete example.
Consider the spacetime manifold $\mathrm{M}$ with metric $\mathrm{g}$ given by
\beq \label{R3S1}
{\mathrm (M ; g)} =
({\mathbb R}^3 \times S^1 ; \eta_{\mu\nu}^{\mathrm Minkowski})\;,
\eeq
and coordinates
\beq
x^0\equiv c\, t,\,x^1,\,x^2\in{\mathbb R} \quad {\mathrm and}\quad x^3 \in [0,L]\;.
\eeq
Take also the chiral gauge field theory with group $G$ and left-handed
fermion representation $R_L$ given by
\beq \label{SO10}
(G \,; R_L) = (SO(10)\, ; {\mathbf 16}+{\mathbf 16}+{\mathbf 16})\;,
\eeq
which incorporates the \SM~(SM) with three families of quarks and leptons.
Then, quantum effects give
\emph{necessarily} CPT violation \cite{K00}, with a typical mass scale
\beq \label{MG}
\msmall_G \equiv \frac{\alpha_G \, \hbar}{L \,c}
  \sim   10^{-35}\: {\rm eV}/c^2 \, \left( \frac{\alpha_G}{0.01}\right)
         \left( \frac{2\: 10^{10}\,{\rm lyr}}{L} \right),
\eeq
where $\alpha_G \equiv g^2/(4\pi)$ is defined in terms of the dimensionless
$SO(10)$ gauge coupling constant $g$ and $L$ is the size of the compact dimension.
As mentioned in the Introduction, this phenomenon has been called the CPT anomaly.
Further discussion of this particular case will be postponed till Sec. IV.
Here, we continue with some general remarks.

\pagebreak
The CPT anomaly occurs for the case of
\beq
SO(10)+ ({\mathbf 16}_L)^3 \quad {\mathrm over}\quad  {\mathbb R}^3 \times S^1
\quad {\mathrm or}\quad  {\mathbb R} \times S^2  \times S^1  \;,
\eeq
but not for
\beq
SO(10)+ ({\mathbf 16}_L)^3 \quad {\mathrm over}\quad  {\mathbb R} \times
S^3\;,
\eeq
where the space manifold $S^3$ is simply connected.
The CPT anomaly also does not occur for the case of
\beq
{\mathrm QED} \quad {\mathrm over}\quad  {\mathbb R}^3 \times S^1 \;,
\eeq
where QED stands for the vector-like gauge theory of
photons and electrons (Quantum Electrodynamics), with $G=U(1)$ and $R_L= 1 + (-1)$.
Hence, \emph{both} topology and parity violation are crucial ingredients
of the CPT anomaly.
(The precise conditions for the occurrence of the CPT anomaly have been given in
Ref. \cite{K00}; see also Sec. IV below.)

As regards the role of topology, the CPT anomaly resem\-bles the Casimir effect,
with the local properties of the vacuum depending on the \bcs; cf. Ref. \cite{DW75}.
Note that the actual topology of our universe is unknown \cite{LR99},
but theoretically there may be some constraints \cite{CW95}.
Interestingly, the modification of the local physics due to the CPT anomaly
(see Sec. V)
would allow for an indirect observation of the global spacetime structure.

Clearly, it is important to
be sure of this surprising effect and to understand the mechanism.
In the next section, we, therefore, turn to a relatively simple theory, the
Abelian chiral gauge theory in two spacetime dimensions (2D).
From now on, we put $\hbar$ $=$ $c$ $=$ $1$, except when stated otherwise.

\section{EXACT RESULT IN 2D}

Consider chiral $U(1)$ gauge theory over the torus $T^2$ $\equiv$ $S^1 \times S^1$,
with a Euclidean metric $g_{\mu\nu}(x)=\delta_{\mu\nu}\equiv\text{diag}(1,1)$.
In order to be specific, take the gauge-invariant model with five left-handed fermions
of charges $(1, 1, 1, 1, -2)$.  Furthermore, impose doubly-periodic \bcs~on
the fermions. The corresponding spin structure is denoted by PP.

The effective action $\Gamma^{1111\overline{2}}_{\mathrm PP}\,[a]$ for the
$U(1)$ gauge field $a_\mu(x)$, which  is defined by the functional integral
\beq  \label{pathint}
\exp\left(-\Gamma^{1111\overline{2}}_{\mathrm PP}\,[a] \right)
 = \int \prod_{f=1}^{5}
   \left( {\mathcal D}\bar{\psi}_{Rf}{\mathcal D}\psi_{Lf}\right)_{\mathrm PP}\;
   \exp\left(-S_{\,\mathrm Weyl}^{\,T^2}
   \left[\bar{\psi}_{Rf},\psi_{Lf},a\right]\right)\,,\;
\eeq
is known \emph{exactly} (see Ref. \cite{IN99} and references therein).
Specifically, the effective action is
given in terms of Riemann theta functions; see Appendix A for details.

It can now be checked explicitly that the CPT transformation
\beq \label{aCPTtransform}
a_\mu(x) \, \rightarrow \, - a_\mu(-x)
\eeq
does \emph{not} leave the effective action invariant:
\beq \label{2dCPTanomaly}
\Gamma^{1111\overline{2}}_{\mathrm PP}\,[a]  \, \rightarrow \,
\Gamma^{1111\overline{2}}_{\mathrm PP}\,[a]+ \,\pi i \pmod{2\pi i}\;.
\eeq
This result  \cite{KN01}, which can also be understood heuristically (see
Appendix A), shows unambiguously the existence of a CPT anomaly in this
particular two-dimensional chiral $U(1)$ gauge theory, the crucial
ingredients being the doubly-periodic (PP) \bcs~and the odd number (here, 5)
of Weyl fermions.

\section{PERTURBATIVE RESULT IN 4D}

Return to 3+1 dimensions. Take, again, the $SO(10)$ chiral gauge theory (\ref{SO10})
with $N_{\mathrm fam}=3$ and the cylindrical spacetime manifold
$\mathrm{M}$ $=$ ${\mathbb R}^3$ $\times$ $S^1$
with metric $g_{\mu\nu}(x)=\eta_{\mu\nu}\equiv\text{diag}(-1,1,1,1)$.
For other possibilities, see Sec. 5 of Ref. \cite{K00}.

The effective action  $\Gamma[A]$, for  $A$ $\in$ ${\mathrm so}(10)$, is,
of course, not known exactly.
But the crucial term has been identified perturbatively:
\beq \label{CSlike}
\Gamma_{\mathrm CS-like}^{\,{\mathbb R}^3 \times S^1}[A\,] =
   \int_{{\mathbb R}^3} \rmd x^0 \rmd x^1 \rmd x^2\, \int_{0}^{L} \rmd x^3
   \; \frac{n \,\pi}{L}
   \: \omega_{\mathrm\, CS}[ A_0(x),A_1(x),A_2(x)\,]\,,
\eeq
with the \CS~density
\beq \label{omegaCS}
\!\!\! \omega_{\mathrm\, CS} [ A_0,A_1,A_2\,] \equiv
       \frac{1}{16 \,\pi^2} \;\epsilon^{3klm}\, {\mathrm tr}
       \Bigl( F_{kl}\, A_m - {\textstyle \frac{2}{ 3}}\, A_k A_l A_m \Bigr),
\eeq
in terms of the \YM~field strength
\beq
F_{kl} \equiv \partial_k A_l - \partial_l A_k + A_k A_l - A_l A_k  \;.
\eeq
Here, the gauge field takes values in the Lie algebra,
$A_m (x)$ $\equiv$ $A_m^a(x) \,T^a$, with normalization
${\mathrm tr}\, (T^a T^b)$ $=$ $(-1/2)\, \delta^{ab}$,
and $\epsilon^{\kappa\lambda\mu\nu}$ is the completely antisymmetric
Levi-Civita symbol, normalized by $\epsilon^{0123}= -\, 1$.          
The Latin indices in Eq. (\ref{omegaCS}) run over $0, 1, 2$, but the fields depend on
\emph{all} coordinates $x^\mu$, for $\mu = 0, 1, 2, 3$.
Note that the effective action term (\ref{CSlike}) is called \CS-like,
because a genuine topological \CS~term exists only in an odd number of
dimensions. At this point we can make three basic observations.

First, the local \CS-like term (\ref{CSlike}) is manifestly Lorentz noninvariant and
CPT-odd, in contrast to the \YM~action,
\beq \label{SYM}
S_{\mathrm\, YM}^{\,{\mathbb R}^3 \times S^1}[A\,] =
     \int_{{\mathbb R}^3} \rmd x^0 \rmd x^1 \rmd x^2\, \int_{0}^{L} \rmd x^3  \;
                   {\textstyle \frac{1}{2}}\,g^{-2}\,
                  {\mathrm tr}\,\Bigl(  \eta^{\kappa\mu}\,\eta^{\lambda\nu} \,
                   F_{\kappa\lambda}(x)\,F_{\mu\nu}(x) \Bigr)\;.
\eeq
More precisely, the Lorentz and CPT transformations considered are active
transformations on fields of local support, as discussed in Sec. IV of Ref.
\cite{KN01}. In physical terms, the wave propagation from the action
(\ref{SYM}) is isotropic, whereas the term (\ref{CSlike}) makes the propagation
anisotropic; see also Sec. V below.
Moreover, both the quadratic and cubic terms in the  \CS-like term  (\ref{CSlike})
can be seen to be T-odd and C- and P-even.
(That both terms transform in the same way under the discrete symmetries
is consistent with the observation that both terms are needed to make the
action (\ref{CSlike}) invariant under infinitesimal non-Abelian gauge
transformations; cf. Sec. 3, p. 239, of Ref. \cite{K98}.)

Second, the integer $n$ in the effective action term (\ref{CSlike})
is a remnant of the ultraviolet regularization:
\beq  \label{n}
n \equiv \sum_{f=1}^{N_{\mathrm fam}} \;(2\, k_{0f}+1)\;,
\quad k_{0f} \in {\mathbb Z}\;,
\eeq
with $N_{\mathrm fam}=3$ for the case considered.
Since the sum of an odd number of odd numbers is odd, one has
\beq \label{nnot0}
n \neq 0 \;,  \quad {\mathrm for} \quad N_{\mathrm fam}=3\;,
\eeq
and the anomalous term (\ref{CSlike}) is necessarily present in the effective action.
For $N_{\mathrm fam}=3$, the regularization of Ref.~\cite{K00} gives simply
\beq \label{nsimplest}
n = (1-1+1)\, \Lambda_0 \, /  \, |\Lambda_0| = \pm \,1 \;,
\eeq
with $\Lambda_0$ an ultraviolet Pauli--Villars
cutoff for the $x^3$-independent modes of the
fermionic fields contributing to the effective action.
The effective action term (\ref{CSlike}) has, therefore,
a rather weak dependence on the small-scale
structure of the theory, as shown by the factor $\Lambda_0/|\Lambda_0|$
in  Eq. (\ref{nsimplest}).
(This weak dependence on the ultraviolet cutoff was first seen in the
so-called ``parity'' anomaly of three-dimensional gauge theories \cite{R84},
which underlies the four-dimensional CPT anomaly discussed here \cite{K98}.)

Third, for the $SO(10)$ theory with \emph{three\,} identical irreducible representations
(irreps), the CPT anomaly \emph{must} occur
[the integer $n$ is odd and therefore nonzero;
cf. Eqs. (\ref{n}) and (\ref{nnot0})$\,$].
For the \SM, the CPT anomaly
\emph{may or may not} occur, depending on the ultraviolet regularization.
The reason is that the SM irreps come in \emph{even} number (for example,
four left-handed isodoublets per family), so that the integer $n$ is not guaranteed to
be nonzero [$n$ is even]. For further details on this subtle point,
see again Sec. 5 of Ref. \cite{K00}.

Next, consider the electromagnetic $U(1)$ gauge field $a_\mu(x)$
embedded in the $SO(10)$ gauge field $A_\mu(x)$
and take $n= - 1$. After the appropriate rescaling of $a_\mu(x)$, the effective
action at low energies then contains the following local terms:
\beqa \label{S-MCS}
S_{\mathrm\, MCS}^{\,{\mathbb R}^3 \times S^1}[a] &=&
 \int_{{\mathbb R}^3} \rmd x^0 \rmd x^1 \rmd x^2\, \int_{0}^{L} \rmd x^3  \;
      \Bigl( {\mathcal L}_{\mathrm\, Maxwell}\,[a]
              + {\mathcal L}_{\mathrm\, CS-like}\,[a]\Bigr)
              \;,      \\[0.25cm]
{\mathcal L}_{\mathrm\, Maxwell}\,[a] &=&
              - \textstyle{\frac{1}{4}}\, \eta^{\kappa\mu}\,\eta^{\lambda\nu} \,
              f_{\kappa\lambda}\,f_{\mu\nu} \;,\label{Max}\\[0.25cm]
{\mathcal L}_{\mathrm\, CS-like}\,[a] &=&
                + {\textstyle \frac{1}{4}}\,\msmall\,
                \epsilon^{3\kappa\lambda\mu} \, f_{\kappa\lambda}\,a_\mu\;
                ,\label{CS}
\eeqa
with the definitions
\beq
f_{\mu\nu} \equiv  \partial_\mu a_\nu   -   \partial_\nu a_\mu  \;,\quad
\msmall          \sim \alpha/L \;, \quad \alpha \equiv e^2/(4\pi) \;.  \nonumber
\eeq
The precise numerical factor in the definition of $\msmall$ depends on the
details of the unification and the running of the coupling constant.

Now focus on the  Maxwell--\CS~(MCS) theory (\ref{S-MCS}) \emph{per se}.
The action $S_{\mathrm\, MCS}^{\,{\mathbb R}^3 \times S^1}$ is gauge
invariant \cite{CFJ90}, provided the electric and magnetic fields in
$f_{kl}$ vanish fast enough as
$(x^0)^2+ (x^1)^2+ (x^2)^2 \rightarrow \infty$. (This observation
makes clear that the parameter $\msmall$ is not sim\-ply the mass
of the photon \cite{GN71}, it affects the propagation in a different way;
cf. Sec. V.)

On the other hand, there is known to be a close relation  \cite{J57,BLOT90} between
CPT invariance and microcausality, i.e., the commutativity of local observables with
spacelike separations. How about causality in the CPT-violating MCS theory?
Remarkably, microcausality (locality)
can be established also in the MCS theory \cite{AK01npb}.
Details are given in Appendix B.

\section{PHENOMENOLOGY: PROPAGATION OF LIGHT}

The propagation of light in the Maxwell--\CS~(MCS) theory (\ref{S-MCS})
makes clear that C and P are conserved, but T \emph{not}.
An example is provided by the behavior of pulses of circularly polarized light,
as will now be discussed briefly.

The dispersion relation for plane electromagnetic waves in the MCS theory
is given by \cite{AK01npb,CFJ90,CK98}
\beq \label{eq:disprel}
\omega_{\,\pm}^2 \equiv k_1^2 + k_2^2 +
                    \bigl( q_3 \pm \msmall/2   \bigr)^2 \; ,
                    \quad q_3 \equiv \sqrt{k_3^2 + \msmall^2 /4} \;,
\eeq
where the suffix $\pm$ labels the two different modes. The phase and group velocities
are readily calculated from this dispersion relation,
\beq \label{eq:vphvg}
\vec v_{\rm ph}^{\, \pm} \equiv \bigl(k_1,k_2,k_3\bigr) \,
                                \frac{\omega_{\, \pm}}{|\vec k\,|^2}    \;, \quad
\vec v_{\rm g}^{\, \pm}  \equiv
\left( \frac{\partial}{\partial k_1}\,, \frac{\partial}{\partial k_2}\,,
       \frac{\partial}{\partial k_3}\,\right)\,
                       \omega_{\,\pm}\; .
\eeq
The magnitudes of the group velocities turn out to be given by
\beq \label{eq:absvg2}
|\vec v_{\rm g}^{\, \pm} (k_1,k_2,k_3)|^2 = \frac
{k_1^2 + k_2^2 +  \bigl(q_3 \pm \msmall/2 \bigr)^2 \,k_3^2 /q_3^2 }
{k_1^2 + k_2^2 +  \bigl(q_3 \pm \msmall/2 \bigr)^2 } \leq 1 \;,
\eeq
with equality for $m=0$ (recall $c \equiv 1$).
Strictly speaking, the wave vector component $k_3$ is discrete
($k_3 = 2\pi n_3/L$, with $n_3 \in \mathbb{Z}$),
but here $k_3$ is considered to be essentially continuous.

For our purpose, it is necessary to give the electric and
magnetic fields of the two modes
in detail. As long as the propagation of the plane wave is
not exactly along the $x^3$ axis, the radiative electric field can be
expanded as follows ($\Re$ denotes taking the real part):
\beq \label{E}
\vec E_{\,\pm} (\vec x,t) =
\Re\,\Biglb(
     c^{\,\pm}_1 \left( \hat e_3 - (\hat e_3 \cdot \hat k) \,\hat k\right) +
     c^{\,\pm}_2 \left( \hat e_3 \times \hat k \right) +
     c^{\,\pm}_3  \, \hat k \, \sin\theta \, \Bigrb) \,
     \exp \left[\, i(\vec k \cdot \vec x - \omega_{\,\pm} \,t) \,\right] ,
\eeq
with unit vector $\hat e_3$ in the compact $x^3$ direction,
unit vector $\hat k$ corresponding to the wave vector $\vec k\,$,
polar angle $\theta$ of the wave vector
(so that $k_3 \equiv \vec k \cdot \hat e_3 = |\vec k|\,\cos\theta$),
and complex coefficients $c^{\,\pm}_1$, $c^{\,\pm}_2$, and $c^{\,\pm}_3$
(at this point, the overall normalization is arbitrary).
The vacuum MCS field equations \cite{CFJ90} then give
the following polarization coefficients for the two modes:
\beq \label{Epols}
\left(\begin{array}{c}
c^{\,\pm}_1 \\[1mm]
c^{\,\pm}_2 \\[1mm]
c^{\,\pm}_3
\end{array}\right) =
\left(\begin{array}{c}
\cos\theta \left(\sqrt{\cos^2\theta +
\mu_{\,\pm}^2 \sin^4\theta} \,\pm\,  \mu_{\,\pm} \,\sin^2\theta\right)^{-1}
\\ \pm \,i \\[1mm]
\mp \,2\, \mu_{\,\pm} \, \sin\theta
\end{array}\,\right)\;, \quad  \mu_{\,\pm} \equiv \frac{m}{2\,\omega_{\,\pm}}\;,
\eeq
with the positive frequencies $\omega_{\,\pm}$ from Eq. (\ref{eq:disprel}).
The corresponding magnetic field is
\beq
\vec B_{\,\pm} = \Re\; \Bigl(\vec k \,\times \vec E_{\,\pm} \Bigr)/\,\omega_{\,\pm}\;.
\eeq

As long as
the $ \mu_{\,\pm} \sin^2\theta$ terms in Eq. (\ref{Epols}) are negligible
compared to $|\cos\theta|$, the transverse electric field
consists of the usual circular polarization modes (see below).
For the opposite case, $|\cos\theta|$ negligible compared to $ \mu_{\,\pm} \sin^2\theta$,
the transverse polarization ($c^{\,\pm}_1$, $c^{\,\pm}_2$) becomes effectively
linear, which agrees with the general remarks in Sec. IV B of Ref. \cite{CK98}.

Now consider the propagation of light pulses close to the $x^2$ axis.
For $k_1=0$ and $0 < \msmall \ll 2\pi/L \ll |k_3| \ll |k_2|$ in particular,
we can identify the $\pm$ modes of the dispersion relation (\ref{eq:disprel})
with left- and right-handed circularly polarized modes
($L$ and $R\,$; see Ref.~\cite{J75}),
depending on the sign of $k_3 \equiv |\vec k|\,\cos\theta$.
From Eqs. (\ref{E}) and (\ref{Epols}), one obtains that
$+/-$ corresponds to $R/L$ for $k_3>0$ and to $L/R$ for $k_3<0$.  

Equation (\ref{eq:absvg2}) then gives the following relations for the group
velocities of pulses of circularly polarized light:
\beqa \label{vgP}
|\,\vec v^{\,L}_{\mathrm g}(0,k_2,k_3)| =  |\,\vec v^{\,R}_{\mathrm g}(0,-k_2,-k_3)|\;,
\eeqa
and
\beqa \label{vgT}
|\,\vec v^{\,L}_{\mathrm g}(0,k_2,k_3)| \neq |\,\vec v^{\,L}_{\mathrm g}(0,-k_2,-k_3)|\;,
\eeqa
as long as $\msmall \neq 0$.
Recall at this point that the time-reversal operator T reverses the
direction of the wave vector and leaves the helicity unchanged,
whereas the parity-reflection operator P flips both  the wave vector and the helicity.
The equality (\ref{vgP}) therefore implies parity invariance and the inequality
(\ref{vgT}) time-reversal noninvariance for this concrete physical situation (see Fig.~1).

Furthermore, the vacuum has become \emph{optically active},
with left- and right-handed monochromatic plane waves traveling at different speeds
\cite{CFJ90}, as follows from Eq. (\ref{eq:disprel}) above (see also Fig.~1).
As mentioned in Ref. \cite{K00}, this may lead to observable effects of the CPT anomaly
in the cosmic microwave background: the polarization pattern
around hot- and cold-spots is modified, due to the
action of the \CS-like term (\ref{CS}) on the  electromagnetic
waves traveling between the last-scattering
surface (redshift $z$ $\sim$ $10^3$) and the detector ($z$ $=$ $0$).
Figure 2 gives a sketch of this cosmic
birefringence, which may be
looked for by NASA's Microwave Anisotropy Probe and ESA's Planck Surveyor.
See Ref. \cite{HW97} for a pedagogical review of the expected \cmb~polarization
and Ref. \cite{L98}
for further details on the possible signatures of cosmic birefringence.

It is important to realize that the optical activity from the CPT anomaly, as
illustrated by Fig.~2, is essentially frequency independent,
in contrast to the quantum
gravity effects suggested by the authors of Ref. \cite{GP99}, for example. In general,
quantum gravity effects on the photon propagation can be expected
to become more and more important as the photon frequency increases towards
$M_{\mathrm \,Planck}\equiv (\hbar\, c/G)^{1/2}\sim 10^{19}\,{\mathrm GeV}$.
The potential CPT anomaly effect at the relatively low frequencies
($\sim 10^{-4}\,{\mathrm eV}$) of the \cmb~is, therefore,
quite remarkable; see also the comments below Eq. (\ref{nsimplest}).

\section{CONCLUSIONS}

The possible influence of the spacetime topology on the local properties of
\qfth~has long been recognized (e.g., the Casimir effect).
As discussed in the present talk, it now appears that nontrivial topology
may also lead to CPT noninvariance for chiral gauge field theories such as
the \SM~with an odd number of families.
This holds even for flat spacetime manifolds, that is,  without gravity.

As to the physical origin of the CPT anomaly, many questions remain (the same
can be said about chiral anomalies in general). It is, however, clear that the
gauge-invariant second-quantized vacuum state plays a crucial role in
connecting the global spacetime structure to the local physics \cite{K00,K98}.
In a way, this is also the case for the Casimir effect \cite{DW75}.
New here is the interplay of parity violation (chiral fermions) and gauge invariance.
Work on this issue is in progress, but progress is slow.

As to the potential applications of the CPT anomaly, we can mention:
\begin{itemize}
\item   the \emph{optical activity of the vacuum\,}
       (leading to a polarization effect for the \cmb; cf. Fig.~2);
\item   the \emph{fundamental arrow-of-time\,}
       (possibly playing a role at the beginning of the universe;
       cf. Refs. \cite{K00,P79}).
\end{itemize}

An important property of the four-dimensional CPT anomaly is the ultraviolet/infrared
connection, exemplified by the factor $n/L$ in the  effective action term
(\ref{CSlike}). Perhaps this allows us to get a handle on the small-scale
structure of spacetime (wormholes, spacetime foam, spin network, \ldots)
by studying the long-range propagation of photons.

\acknowledgments

It is a pleasure to thank the organizers of the conference for their hospitality,
J. Pullin for correspondence, C. Adam for help with the signs,
and C. Mayer for help with the figures.

\appendix

\section{EFFECTIVE ACTION FOR 2D CHIRAL $U(1)$ GAUGE THEORY}

The two-dimensional Euclidean action for a single one-component Weyl field $\psi(x)$ of
unit electric charge over the particular torus $T^2$
with modulus $\tau=i$ is given by
\beq \label{eq:action}
S_{\,\mathrm Weyl}^{\,T^2}\left[\bar{\psi},\psi,a\right]
 =                - \int_{0}^{L} \rmd  x^1 \int_{0}^{L}\rmd  x^2 \;e\: \bar{\psi}\:
                    e^\mu_a\,\sigma^a \!\left(\partial _\mu + i a_\mu  \right)  \psi \;,
\eeq
with
\beq
   (\sigma^1, \sigma^2) = (1,i)\,, \;\;
   e^\mu_a = \delta^\mu_a\,\,, \;\; e \equiv \det\left( e_\mu^a\right) = 1 \,.
\eeq
The $U(1)$ gauge potential can be decomposed as follows:
\beq
a_\mu (x)= \epsilon_{\mu\nu}\, \delta^{\nu\rho}\,\partial _\rho \phi (x)  +
           2\pi h_\mu/L + \, \partial_\mu \chi (x)  \;,
\label{eq:Adecomposed}
\eeq
with $\phi(x)$ and $\chi(x)$  real periodic functions
and  $h_1$ and $h_2$ real constants.
Here, $\chi(x)$ corresponds to the gauge degree of freedom.
The related gauge transformations on the fermion fields are
\beq
\psi(x)       \rightarrow \exp[-i\chi(x)\,]\;\psi(x) \;, \quad
\bar{\psi}(x) \rightarrow \exp[+i\chi(x)\,]\;\bar{\psi}(x) \;.
\eeq

Next, impose doubly-periodic \bcs~on the fermions,
\beq
   \psi  (x^1 + L, x^2) =  +\,\psi  (x^1,x^2)\;, \quad
   \psi  (x^1, x^2 + L) =  +\,\psi  (x^1,x^2)\;.
\eeq
This spin structure will be denoted by
PP, where P stands for periodic \bcs.
(The other spin structures are AA, AP, and PA, where A stands for antiperiodic \bcs.)

The effective action $\Gamma_{\mathrm PP}\,[a]$ of this $(1111\overline{2})$-model,
defined by the functional integral (\ref{pathint}), is found to be given by
\cite{IN99}
\beq \label{GammaPP}
\exp\left(-\Gamma^{1111\overline{2}}_{\mathrm PP}\,[a] \right) \equiv
D^{1111\overline{2}}_{\mathrm PP}\,[a]
  = \left( D_{\mathrm PP}\,[a]  \right)^4 \,
                       \overline{\left( D_{\mathrm PP}[2a] \right)}\;,
\eeq
with the single chiral determinant
\beqa
\label{eq:DPP}
D_{\mathrm PP}\,[a] &=&
\hat{\vartheta} (h_1 + {\textstyle\frac{1}{2}},
                                   h_2 + {\textstyle\frac{1}{2}} )\;
                 \exp\left(i\,\frac{\pi}{2}\,(h _ 1 - h _2) \right)\;
\exp\left(\frac{1}{4\pi}\int_{T^2} \rmd ^2 x
     \left( \phi \,\partial ^2 \phi + i \phi \,\partial ^2
     \chi\right)\right) .
\eeqa
Here, the complex function
\beqa
\label{eq:thetahat}
\hat{\vartheta} (x,y) &\equiv& \exp\left(- \pi y^2 + i \pi x y\right)
                                    \, \vartheta (x+i y;i)/ \eta (i)  \; ,
\quad {\mathrm for}\quad  x,y \in {\mathbb R} \;,
\eeqa
is defined
in terms of the Riemann theta function $\vartheta (z;\tau)$
and De\-de\-kind eta function $\eta (\tau )$, for modulus $\tau=i$.
The bar on the \rhs~of Eq.~(\ref{GammaPP}) denotes complex conjugation.

The gauge invariance of the effective action (\ref{GammaPP}) can be readily verified.
In fact, the gauge degree of freedom $\chi(x)$ appears only in the exponential
of Eq. (\ref{eq:DPP}),
namely in the term proportional to $i \phi \,\partial ^2 \chi$, and cancels out for
the full expression (\ref{GammaPP}), since $4 \times 1^2$ $-$ $1 \times 2^2$ $=0$.
The invariance  under large gauge transformations
$h_\mu$ $\rightarrow$ $h_\mu+n_\mu$, for $n_\mu \in {\mathbb Z}$,
requires a little bit more work.

The CPT anomaly (\ref{2dCPTanomaly}) follows directly
from the $\vartheta$ function properties \cite{KN01}.
The relevant properties of $\vartheta (z;\tau)$
are its periodicity under $z$ $\rightarrow$ $z+1$ and
quasi-periodicity under $z$ $\rightarrow$ $z+\tau$, together with the
symmetry $\vartheta (-z;\tau)$ $=$ $\vartheta (z;\tau)$.
But the anomaly can also be understood heuristically from the product of eigenvalues.
For gauge fields (\ref{eq:Adecomposed}) with
$\phi(x)$ $=$ $\chi(x)$ $=$ $0$
and infinitesimal harmonic pieces $h_\mu$, one has, in fact,
\beq \label{eq:D11112linear}
D^{1111\overline{2}}_{\mathrm PP}\,[h_1,h_2] =
c\,(h_1 + i h_2)^3\,(h_1^2 + h_2^2) + {\rm O}(h^7) \;,
\eeq
with a nonvanishing complex constant $c$. Clearly, this expression
changes sign under the transformation
$h_\mu \rightarrow - h_\mu$, which corresponds to the CPT transformation
(\ref{aCPTtransform}). See Ref. \cite{KN01} for further details.

By choosing topologically nontrivial zweibeins $e_\mu^a(x)$
[still with a flat metric
$g_{\mu\nu}(x)\equiv  e_\mu^a(x)\;e_\nu^b(x)\;\delta_{ab}=\delta_{\mu\nu}\,$]
and including the spin connection term in the covariant derivative of the
fermionic action (\ref{eq:action}),
the CPT anomaly can be moved to the spin structures AA, AP, and PA.
These topologically nontrivial zweibeins
correspond, however, to the presence of spacetime torsion, which may be of
interest in itself. See Ref. \cite{KM01} for further details on role of
topologically nontrivial torsion.

\section{MICROCAUSALITY IN 4D MCS THEORY}

For the Maxwell--\CS~(MCS) theory (\ref{S-MCS})
in  the Coulomb gauge $\nabla \cdot \vec a=0$ (with vector indices
running over 1, 2, 3, and  $\hbar$ $=$ $c$ $=$ $1$),
the following commutators have been found for the electric field
$\vec e \equiv \partial_0 \,\vec a -\nabla\, a_0$ and magnetic field
$\vec b \equiv \nabla \times \vec a$ :
\beqa
[e_i (x),e_j (0)]&=&
\Biglb( (\delta_{ij}\,\partial_0^2 -\partial_i \partial_j )\,(\partial_0^2 -\nabla^2 )
+ \msmall^2 \,\delta^3_i \delta^3_j \, \partial_0^2
 \no \\[0.1cm]
&&
- \,\msmall\, \epsilon_{ij3}\, \partial_0^3  + \msmall\,
( \epsilon_{ia3} \,\partial_j  - \epsilon_{ja3} \,\partial_i  )\,\partial_a \partial_0
\Bigrb)  \, i D_{\mathrm\, MCS}(x) \,,\label{eq:comm-ee}
\\[0.2cm]
[e_i (x),b_j (0)] &=&
\Biglb( \epsilon_{ijl}\, \partial_l\partial_0 \,(\partial_0^2 -\nabla^2 )
 - \msmall^2 \, \delta^3_i\, \epsilon_{j3a}\, \partial_a \partial_0
\no \\[0.1cm]
&&
+ \,\msmall  \, \left( \delta_{ij}\,\partial_0^2  - \partial_i \partial_j\right) \partial_3
-\msmall \,  \delta^3_j \,\partial_i (\partial_0^2 - \nabla^2 )
\Bigrb) \,i D_{\mathrm\, MCS}(x)\; ,\label{eq:comm-eb}
\\[0.2cm]
[b_i (x), b_j (0)] &=&
\Biglb( (\delta_{ij}\nabla^2 -\partial_i \partial_j )\,(\partial_0^2 -\nabla^2 )
-\msmall \,\epsilon_{ijl}\, \partial_l \partial_0 \partial_3
\no \\[0.1cm]
&&
+\,\msmall^2 \,\Bigl( \delta_{ij}\,(\nabla^2 - \partial_3^2 )
- \partial_i\partial_j  - \delta^3_i  \delta^3_j\, \nabla^2
+( \delta^3_i\, \partial_j  + \delta^3_j\,\partial_i  )\, \partial_3 \Bigr)
\Bigrb) \,i D_{\mathrm\, MCS}(x) \;, \label{eq:comm-bb}
\eeqa
with the commutator function
\beqa \label{eq:comm-func-space}
D_{\mathrm\, MCS}(x)&\equiv&(2\pi)^{-4}\,\oint_C \rmd p_0 \int \rmd^3 p\;
        \frac{  \exp \left[\,ip_0\, x^0 +i\vec p \cdot \vec x \,\right]  }
             {\left(|\vec p\,|^2  -p_0^2 \right)^2 +
                \msmall^2 \left(p_1^2 + p_2^2 -p_0^2 \right)} \;\;,
\eeqa
for a contour $C$ that encircles all four poles of the
integrand in the counterclockwise direction.
Note that the derivatives on the \rhs~of Eqs. (\ref{eq:comm-ee})--(\ref{eq:comm-bb})
effectively bring down powers of the momenta in the integrand of Eq.
(\ref{eq:comm-func-space}).
[The calculation of the commutators (\ref{eq:comm-ee})--(\ref{eq:comm-bb})
is rather subtle: $a_0$, for example, does not vanish in the Coulomb gauge
but is determined by a nontrivial equation of motion,
$a_0 =i\,\msmall \,|\vec p\,|^{-2}\,\epsilon_{3kl}\,a_k \,p_l $ in momentum space.]

The Lorentz noninvariance of the MCS theory is illustrated by the denominator of the
integrand in Eq. (\ref{eq:comm-func-space}) and the fact that
commutators (\ref{eq:comm-ee}) and (\ref{eq:comm-bb}) differ at order $\msmall^2$.
Two further observations can be made:
\begin{enumerate}
\item The commutator function vanishes for spacelike separations,
\begin{equation}
D_{\mathrm\, MCS}(x^0 ,\vec{x}\,)=0\,, \quad {\mathrm for}\;\; |x^0 |<|\vec{x}\,| \; ,
\end{equation}
as follows by direct calculation.
\item Even though the commutators of the vector potentials $ \vec a\, (x)$
      have poles of the type $|\vec p\,|^{-2}$, these poles, which spoil causality,
      are absent for the commutators (\ref{eq:comm-ee})--(\ref{eq:comm-bb})
      of the physical (gauge-invariant) electric and magnetic fields.
\end{enumerate}
Hence, the locality results of QED~\cite{JP28} carry over to the
MCS theory, at least for the ``space\-like''
\CS~term  (\ref{CS}) considered \cite{AK01npb}.
This MCS theory has, however, new uncertainty relations
(e.g., for the $b_1$ and $b_2$ fields averaged over the same spacetime region).

The ``timelike'' MCS theory, with $ \epsilon^{3\kappa\lambda\mu}$
in (\ref{CS}) replaced by $ \epsilon^{0\kappa\lambda\mu}$, does
violate microcausality, as long as unitarity is enforced \cite{AK01npb}.
This particular result may have other implications.
It rules out, for example, the possibility that a \CS-like
term can be radiatively induced from a CPT-violating
axial-vector term in the Dirac sector \cite{AK01plb}.

\begin{figure}\vspace{1cm}
\centerline{\psfig{figure=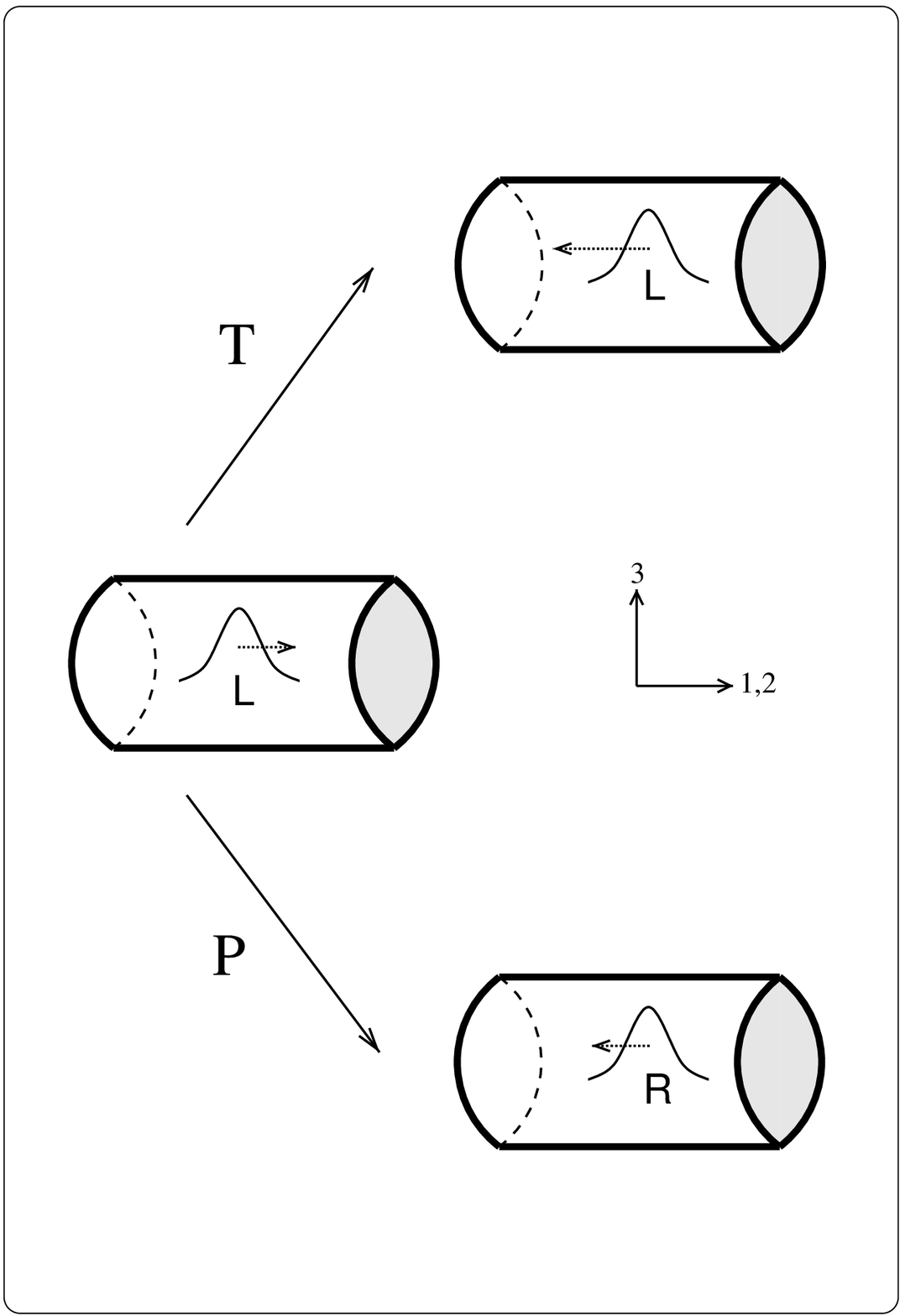,width=0.6\hsize}}\vspace{3\baselineskip}
\caption{Sketch of the behavior of a left-handed wave packet in the \MCS~theory
  (\ref{S-MCS}) under the time reversal (T) and parity (P) transformations.
  [The charge conjugation (C) transformation acts trivially.] The dotted arrows
  indicate the group velocity approximately in the $x^1$ or $x^2$ direction,
  for the case of a compact $x^3$ coordinate.
  The magnitude of the group velocity changes under T, but not under C or P
  (hence, the physics is CPT-noninvariant).
  In addition, the vacuum is seen to be optically active,
  with left- and right-handed light pulses traveling to the left at different speeds
  (because of parity invariance, the same holds for pulses of circularly polarized
  light traveling to the right).}
\label{fig1}
\end{figure}

\begin{figure}
\centerline{\psfig{figure=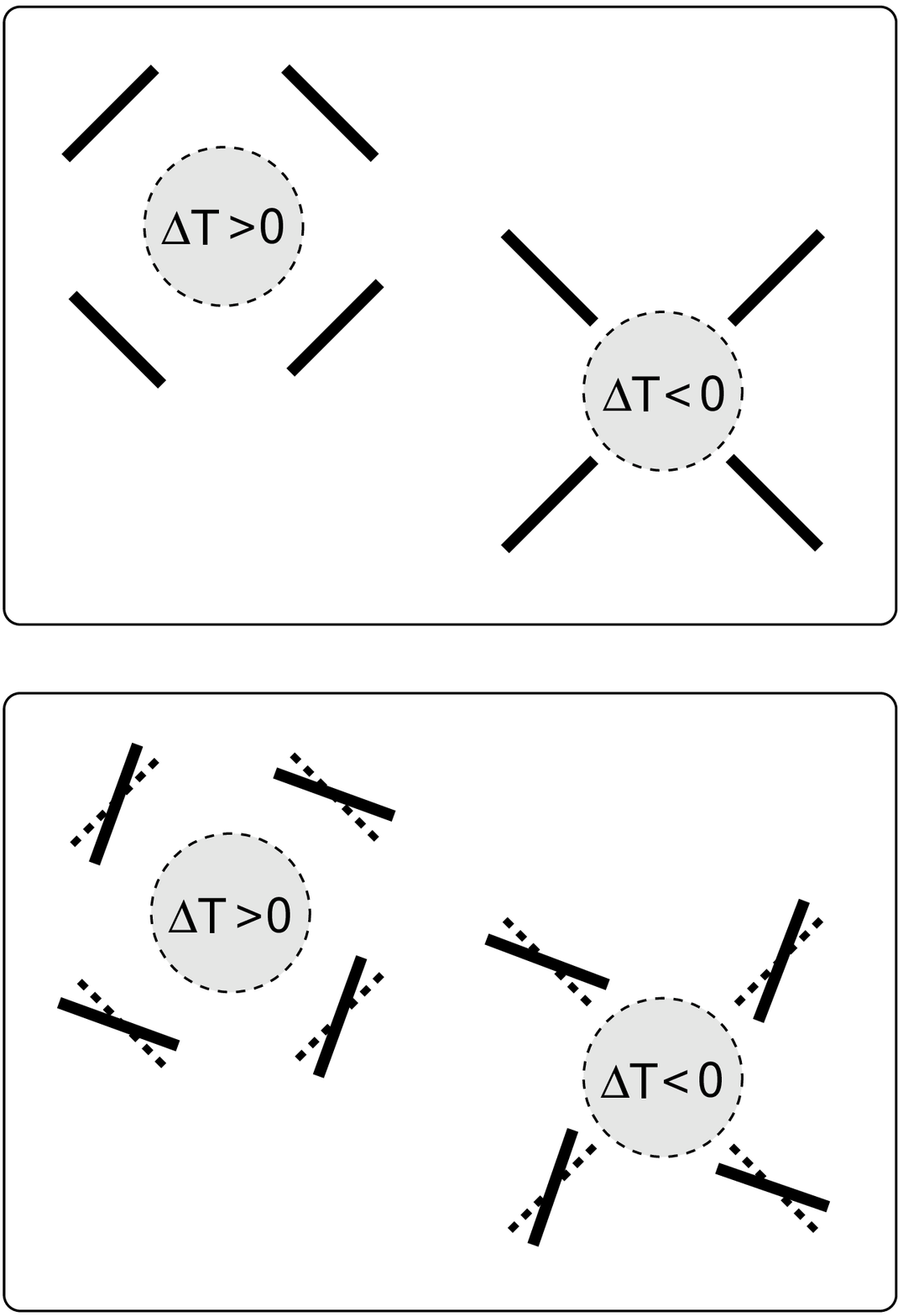,width=0.6\hsize}}\vspace{3\baselineskip}
\caption{Sketch of the linear polarization pattern (indicated by heavy bars)
      around \cmb~hot- and cold
      spots, generated by scalar perturbations of the metric.
      The top panel is in a noncompact direction (corresponding to the $x^1$
      coordinate, say). The bottom panel is in the compact direction
      (corresponding to the $x^3$ coordinate) and displays the optical activity of
      the \MCS~theory (\ref{S-MCS}) considered. In fact,
      for a patch of the sky in a particular direction along the  $x^3$ axis
      (bottom panel), the linear
      polarization pattern is rotated by a very small amount in
      counterclockwise direction.
      For a patch of the sky in the opposite direction (not shown),
      the rotation of the linear polarization is in the clockwise direction.
      Precisely which  particular direction along the  $x^3$ axis corresponds
      to the counterclockwise rotation and which to the clockwise rotation
      depends on the small-scale structure of the theory, that is, the sign
      of the parameter $n$ in the  \CS-like term (\ref{CSlike})
      of the effective action.}
\label{fig2}
\end{figure}

\end{document}